\documentclass[%
 reprint,
 superscriptaddress,
 amsmath,amssymb,
 aps,
 prx,
 longbibliography,
]{revtex4-2}
\usepackage{graphicx}
\usepackage{dcolumn}
\usepackage{bm}
\usepackage{bbm}
\usepackage{hyperref}

\usepackage{xcolor}

\newcommand{\tx}{\underline{x}}
\newcommand{\txx}{\mathbf{\underline{x}}}
\newcommand{\txN}[1]{\mathbf{\underline{x}}_{\partial #1}}
\newcommand{\txeN}[2]{\mathbf{\underline{x}}_{\partial #1 \setminus #2}}

\newcommand{\ty}{\underline{y}}
\newcommand{\tyy}{\mathbf{\underline{y}}}

\newcommand{\Ncl}{\mathcal{N}}

\newcommand{\xx}{\mathbf{x}}

\newcommand{\R}{{\mathbb R}}

\newcommand{\N}{{\mathbb N}}

\begin{document}

\preprint{APS/123-QED}

\title{Backtracking Dynamical Cavity Method}

\author{Freya Behrens}
\affiliation{%
 Statistical Physics Of Computation Laboratory, \'Ecole Polytechnique F\'ed\'erale de Lausanne, Lausanne, Switzerland
}%
  \author{Barbora Hudcová}%
  \affiliation{%
Algebra Department, Faculty of Mathematics and Physics,  Charles University, Prague, Czech Republic
}%
 \affiliation{%
Czech Institute of Informatics, Robotics and Cybernetics, Czech Technical University, Prague, Czech Republic
}%
\author{Lenka Zdeborová}%
\affiliation{%
 Statistical Physics Of Computation Laboratory, \'Ecole Polytechnique F\'ed\'erale de Lausanne, Lausanne, Switzerland
}%

\begin{abstract}
The cavity method is one of the cornerstones of the statistical physics of disordered systems such as spin glasses and other complex systems. It is able to analytically and asymptotically exactly describe the equilibrium properties of a broad range of models. Exact solutions for dynamical, out-of-equilibrium properties of disordered systems are traditionally much harder to obtain. Even very basic questions such as the limiting energy of a fast quench are so far open. 
The dynamical cavity method partly fills this gap by considering short trajectories and leveraging the static cavity method. However, being limited to a couple of steps forward from the initialization it typically does not capture dynamical properties { related to attractors of the dynamics.} 
We introduce the backtracking dynamical cavity method that instead of analysing the trajectory forward from initialization, analyses trajectories that are found by tracking them backward from attractors.
We illustrate that this rather elementary twist on the dynamical cavity method leads to new insight into some of the very basic questions about the dynamics of complex disordered systems. This method is as versatile as the cavity method itself and we hence anticipate that our paper will open many avenues for future research of dynamical, out-of-equilibrium, properties in complex systems. 
\end{abstract}

\maketitle

\section{Introduction.}
The cavity method is one of the main analysis tools to investigate equilibrium properties of disordered and complex systems. It has been introduced in a series of seminal works as an alternative to the replica method for mean-field models of spin glasses \cite{mezard1987spin}. 
Subsequent key developments on diluted lattices aka sparse random graphs 
\cite{mezard2001bethe,mezard2003cavity} and the link between the cavity method and message passing algorithms \cite{kabashima1998belief,mezard2002analytic} have led to an explosion of applications of the method in systems on sparse random structures, such as error correcting codes, random constraints satisfaction problems, random graphs colouring, or community detection to mention just a few of many, see e.g.~the textbook \cite{mezardInformationPhysicsComputation2009}. Results obtained using the cavity method are in many cases exact in the thermodynamic limit which is particularly appealing for theoretical studies in computer science and mathematics. 

Many questions about complex systems of current interest are, however, not concerned with equilibrium properties but with dynamical, out-of-equilibrium, ones. An exact analysis of dynamical properties is much more challenging compared to the equilibrium ones. 
Let us give two concrete examples of very basic questions about dynamics that are so far open and that the method proposed in this paper resolves.

\paragraph*{Example 1:} Consider the anti-ferromagnetic Ising model or a spin glass with random $\pm 1$ interactions on a random $d$-regular graph of $n$ nodes.
Consider then the dynamics where at each time step every spin aligns with their magnetic field or remains in case the field is zero. We initialize each spin randomly. To which value of energy does such a dynamics converge at large times when $n\to \infty$?

\paragraph*{Example 2:} Consider now the ferromagnetic Ising model on a random $d$-regular graph, the same dynamical process but initialized at magnetization $-1 < m < 1$. For what values of $m\ge 0$ does the dynamics go to the homogeneous all $+1$ configuration and for what values of $m \ge 0$ does it go elsewhere when $n\to \infty$? What other attractor types does the dynamics converge to for other values of $m$? 

While these are very basic questions that could be studied numerically in an undergraduate class on statistical mechanics, 
{ the asymptotically exact answer is so far not known} even for random graphs for which many static properties are known exactly in the thermodynamic limit via the cavity method \cite{mezard1987spin,mezard2001bethe,mezard2003cavity}. 

The main contribution of this paper is to present a method to answer dynamical questions such as the above { by quantifying the basin of attraction of different types of attractors for deterministic dynamics}. We call it the \textit{backtracking dynamical cavity method} (BDCM). This method provides a solution in the sense that for models on random graphs in the limit $n\to \infty$ it gives a closed-form analytical prescription of how to compute the desired values. 
This leads for instance, to the value of the limiting energy 
from Example 1 for random regular graphs and to exhibiting different types of attractors and dynamical phase transitions between them in Example 2. 

The main idea behind the backtracking dynamical cavity method is simple. We start with the established idea of the dynamical cavity method {(DCM) \cite{hatchett2004parallel,neriCavityApproachParallel2009,mimuraParallelDynamicsDisordered2009,kanoriaMajorityDynamicsTrees2011,lokhovDynamicMessagepassingEquations2015}} that considers the trajectory of a spin for a finite number of time steps $T$. It considers this trajectory as an augmented $T$-dimensional spin variable and applies the traditional static cavity method to this trajectory-variable. 
The dynamical cavity method provides an exact description of the dynamics as long as the system is large $n\to \infty$ and the time $T=O(1)$ finite. Evaluating the corresponding equations is in general exponentially costly in $T$ and thus limits the choice of $T$. { Consequently, properties that require not-so-small values of $T$ cannot be analyzed using this method unless one resolves to approximations.}

The key twist in the \textit{backtracking} dynamical cavity method is that instead of taking $T$ steps from the beginning of the trajectory, we take $T$ steps from the attractor (thus tracking the dynamics back). { This way we can access properties of the attractors and their basins of attraction $T$ steps back in an asymptotically exact manner without further approximations. We will show that by exploring the last $T=O(1)$ steps of the dynamics the backtracking dynamical cavity method is able to provide answers to the two examples posed above.
What came as a surprise to us, is that }
looking at $T$ step backward covers a basin of attraction of entropy (logarithm of the number of configurations in that basin) very close to the total entropy of all initial conditions already for very moderate values of $T$. 

While the existing DCM is able to access properties that happen in the first few steps of the dynamics \cite{hatchett2004parallel,neriCavityApproachParallel2009,mimuraParallelDynamicsDisordered2009,kanoriaMajorityDynamicsTrees2011,lokhovDynamicMessagepassingEquations2015}, { and with approximations is also able to describe qualitatively correctly large time behavior  even for local observables \cite{aurellDynamicMeanfieldCavity2012,zhangInferenceKineticIsing2012,delferraroDynamicMessagepassingApproach2015,barthelMatrixProductAlgorithm2018,torrisiUncoveringNonequilibriumStationary2022,hurryDynamicsSparseBoolean2022a},} it does not provide asymptotically exact results about the attractors of the dynamics nor their basin of attraction. BDCM does exactly that, describing the last steps of the dynamics. Moreover, as we will see on examples below,
only a few steps back into the basin of attraction may already exhibit qualitative properties of the complete basin of attraction.
We illustrate this in particular on the majority rule where the types of attractors found from initial configurations with different magnetizations already show when we step into the basin of attraction by only one step.




Finally, we want to emphasize that the idea of looking at the last $T$ steps of the dynamics rather than the first $T$ steps is very generic { and open questions about the properties of attractors are abundant in the study of dynamics of complex systems. We thus anticipate that the BDCM will become one of the key analytical methods in the field.}
Possible applications include training dynamics of artificial neural networks { where we would want to study the basin of attraction of a region with good generalization properties; social dynamics on networks where we may want to know what type of Nash equilibria will be reached; gene regulatory networks where attractors correspond to cell types; or various types of far-from-equilibrium physical systems where different attractors may correspond to different phases.} The backtracking idea can be applied not only in conjunction with the dynamical cavity method but  also, for instance, within the dynamical mean field theory \cite{georges1996dynamical} that has been influential in the study of strongly correlated electron systems or neural networks.

\section{Setting and Notation}

By an \emph{undirected graph} of size $n$ we understand the tuple $G = (V, E)$ where $V=\{1, \ldots, n \}$ is the set of nodes and $E=\{(i, j) \, | \, i, \, j \in V \}$ is the set of edges. For each node $i \in V$ we define the \emph{neighbourhood of $i$} to be the set $\partial_i = \{j \, | \, (i, j) \in E \} \subseteq V$ with the degree of $i$ as $d(i)=|\partial_i|$. We say a graph is \emph{$d$-regular} if each node has degree $d$.
Each node $i$ of the graph $G$ can be assigned in one of the discrete states in a set $S$, $x_i \in S$. Such an assignment then represents a configuration $\xx = x_1 \ldots x_n \in S^n$. 
By $\xx_{\partial i}$ we mean the subset of the configuration that includes all neighbours of node $i$.

We consider time-discrete dynamical processes operating on configurations of a graph $G$ with $n$ nodes. 
The state of each node gets updated \textit{synchronously}, the update depends on the node's own state and the state of its neighbours. The dynamical rule is specified for each node individually using the local update function $f_i: S^{1+|\partial i|} \rightarrow S$. This gives rise to a global mapping $F: S^n \rightarrow S^n$ governing the dynamics of the system. For a configuration $\xx \in S^n$, the $i$-th node with neighbourhood $\partial_i = (i_1, \ldots, i_{d(i)})$ gets updated according to 
$$[F(\xx)]_i = f_i(x_i, x_{i_1}, \ldots, x_{i_{d(i)}}).$$ 

To describe the global dynamics, the symbol $\txx$ denotes a sequence of configurations from $S^n$; i.e., $\txx = (\xx^1,...,\xx^t)$ for some $t \in \N$. 
We define the \emph{configuration graph} as an oriented graph whose nodes are the configurations from $S^n$ with edges of the form $(\xx, F(\xx)), \, \xx \in S^n$. 
If $\txx$ satisfies that $\xx^{i+1}=F(\xx^i)$ for each $i$ we call it the \emph{trajectory of length $t$ starting from the initial configuration $\xx^1$}. Since the configuration space is finite, each long enough trajectory becomes eventually periodic. We call the pre-period of the sequence the \emph{transient} and its periodic part the \emph{attractor} or \emph{limit cycle}. 
For an attractor, the set of all configurations converging to it is called its \emph{basin of attraction}. 

In this paper, we will consider the majority dynamics in models with Ising variables and random $\pm 1$ edge weights (covering the Ising ferromagnet, antiferromagnetic and a spin glass). 
Such a dynamics has attractors of length $c \in \{1,2\}$, which is due to an elegant argument on decreasing energy functions by \citep{decreasing-energy_functions,derrida1989dynamical}. The number of attractors and short limit cycles for closely related models have been studied e.g. in \citep{bray1981metastable,behrensDisAssortativePartitions2022a,hwangNumberLimitCycles2020}. Their basin of attraction has, as far as we know, not been studied analytically and we use this as an example of applications of the backtracking dynamical cavity method developed in this paper. 

\section{Backtracking Dynamical Cavity}

\paragraph{General Idea.} The key idea of the backtracking dynamical cavity method (BDCM) is the fact that it acts on static objects that track the dynamics backward from the attractors instead of forward from arbitrary initial states.
To formalize this, we define a $(p/c)$ \emph{backtracking attractor} to be a trajectory of length $p$ that leads into a limit cycle of length $c$ on the configuration graph.
As we increase the length of the incoming trajectory $p$, such an analysis incorporates a growing fraction of an attractor's basin and will illuminate important dynamical questions.

\paragraph{The distribution of backtracking attractors.} For a given global update rule $F$, path length $p$ and cycle size $c$, our goal is to analyze the properties of the attractors and their transients. 
To do this, we introduce a probability distribution over all sequences of configurations $
    \txx=(\xx^1, \ldots, \xx^p, \xx^{p+1} \ldots, \xx^{p+c}) \in (S^n)^{p+c}$ as follows
\begin{eqnarray}
    P(\txx) \!= \!\frac{1}{Z}  \mathbbm{1}\left[F(\xx^{p+c}) = \xx^{p+1}\right]\! \prod_{t=1}^{p+c-1}  \mathbbm{1}\left[F(\xx^{t}) = \xx^{t+1}\right]. \label{eq:prob_dist}
\end{eqnarray}
Here, ${\mathbbm{1}}(\cdot)$ is the indicator function which is $1$ if the Boolean statement is true and $0$ otherwise; $Z$ is the normalization constant of the probability distribution.

A sequence $\txx$ has only non-zero measure if it is consistent with the time evolution of the global update rule due to the term $\prod_{t=1}^{p+c-1} \mathbbm{1}\left[F(\xx^t) = \xx^{t+1}\right]$.
The boundary condition $F(\xx^{p+c}) = \xx^{p+1}$ ensures that this trajectory of configurations ends up in a limit cycle of length~$c$.
Consequently, only $(p/c)$ backtracking attractors can have a non-zero measure in the distribution \eqref{eq:prob_dist}.

Analogous to the classical cavity method for static analysis, the goal is then to compute the free entropy density $\Phi = \frac{1}{n}\log(Z)$, i.e. the logarithm of the number of sequences that are valid backtracking attractors.
Then, $\Phi$ can be viewed as a proxy for the size of an attractor's basin.

\paragraph{Adding observables.}
A key vitrue of the BDCM is that we can obtain this entropy $\Phi$ conditioned on backtracking attractors with specific properties, e.g. fixed energy or magnetization or magnetization in the attractor. 
This can be achieved by flexibly weighting the sequences $\txx$ in the probability distribution according to the relevant observable.
Concretely, one adds the factor
\begin{eqnarray}
    e^{-\sum_{k} \lambda_k \Xi_k(\txx)} \label{eq:extra-factor}
\end{eqnarray}
on the right-hand side of \eqref{eq:prob_dist} and adjusts the normalization $Z$ accordingly; exactly $K$ observables $\Xi_k(\txx)$ are added as summary statistics of the backtracking attractors.
Each observable has an associated parameter $\lambda_k \in \R$ which acts as a temperature from a physics perspective, or as a Lagrangian multiplier viewing the observables as optimization constraints.
We use the notation uppercase $\Xi_k(\txx)$ for the function acting on a trajectory $\txx$.
{The notation lowercase $\xi_k$ is used when the value of $\Xi_k(\txx)/n$ is fixed to $\xi_k$, usually as a constraint and intensive quantity.}
Formally, we define the number of valid backtracking attractors conditioned on fixed observables as $\Ncl(\xi_1,\dots, \xi_K) = e^{n s(\xi_1,...,\xi_K)}$ so that $s$ is their entropy.
Then, the following relation between the entropy $s$ and the normalization constant $Z$ including the extra factor \eqref{eq:extra-factor} holds 
\begin{eqnarray}
     Z &= e^{n\Phi(\lambda_1,\dots,\lambda_K)} = \int_{\txx \in B} e^{- \sum_k \lambda_k \Xi_k(\txx)} \\
    &=\int\left[\prod_k d\xi_k\right] e^{n {[}s(\xi_1,\dots, \xi_K)-\sum_k\lambda_k\xi_k{]}}\,,
\end{eqnarray}
where the set $B$ is the set of all valid $(p/c)$ backtracking attractors.
In the large system limit, when $n \to \infty$, applying the saddle point method on the right-hand side gives an explicit form of the entropy 
\begin{eqnarray}
      s(\hat{\xi}_1,\dots,\hat{\xi}_K)  = \Phi(\lambda_1,\dots,\lambda_K)  + \sum_k \lambda_k \hat{\xi}_k \label{eq:configs}
\end{eqnarray}
under the condition that for all $k=1, \dots, K$
\begin{eqnarray}
     \frac{\partial s(\xi_1,\dots,\xi_K)}{\partial \xi_k}|_{\xi_k=\hat{\xi}_k} = \lambda_k ;\, \\
   \frac{\partial \Phi(\lambda_1,\dots,\lambda_K)}{\partial \lambda_k} = - {\hat{\xi_k} = -\frac{1}{n}\langle \Xi_k \rangle}\,. \label{eq:constraints}
\end{eqnarray}
The $\langle\cdot\rangle$ brackets define an average over the probability measure induced by \eqref{eq:prob_dist}.
As it is infeasible to directly compute $Z$ and $\Phi$ due to the high-dimensional integral over $\txx \in \R^{(p+c)\times n}$ when $n \to \infty$, we  compute the leading order (in $n$) of the free entropy using the replica symmetric cavity method or equivalently belief propagation~\cite{mezardInformationPhysicsComputation2009}.

\begin{figure*}[t]
    \centering
 \includegraphics[height=0.22\linewidth]{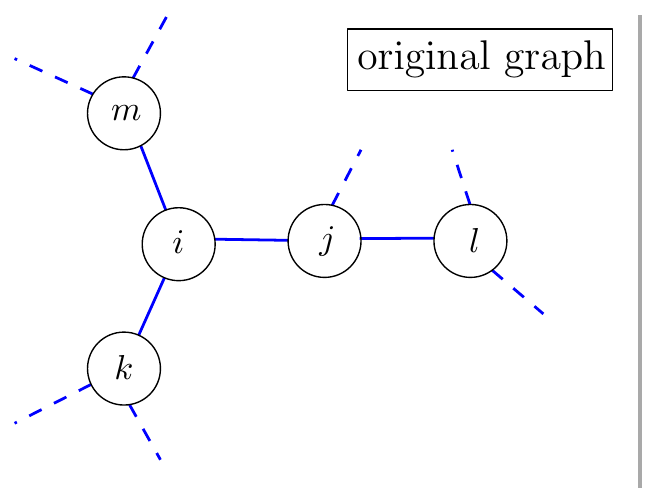}
\includegraphics[height=0.22\linewidth]{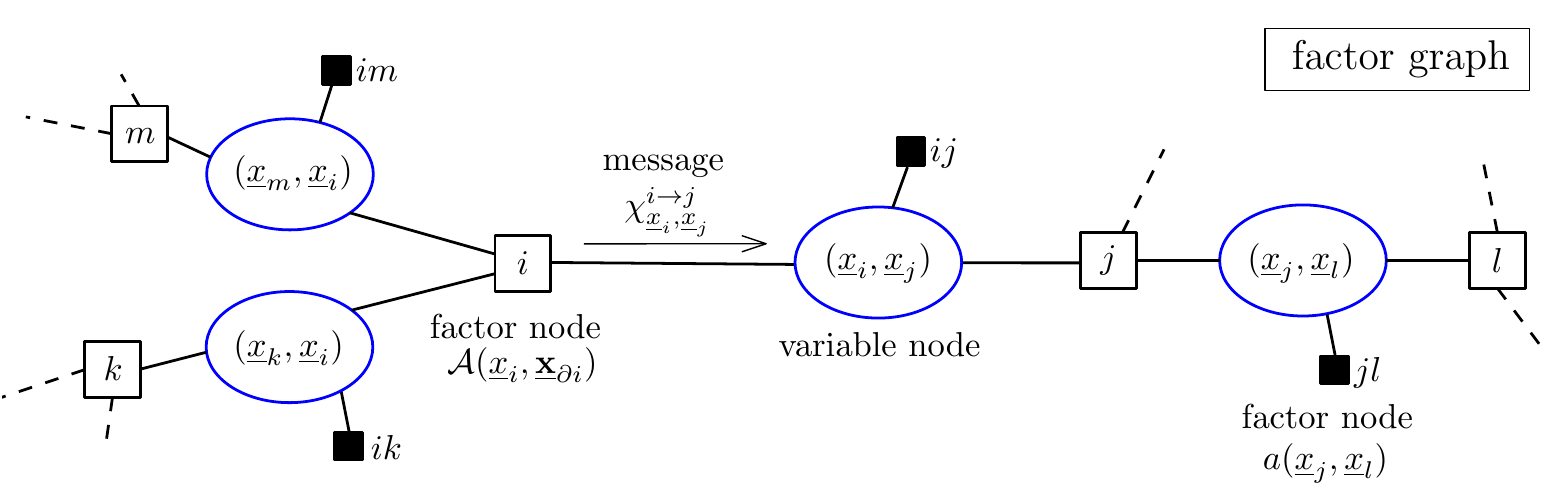}
    \caption{\textbf{A subgraph of the factor graph for the BDCM on a $3$-regular graph.} \textit{(Left)} Original graph. \textit{(Right)} Factor graph in the edge dual space. 
    The tuples in the round variable nodes  can take on values of all possible trajectories $\tx$. 
    The factor nodes on tuples can enforce the constraints on the variables on their own.
    The factor~nodes between $d$ tuples correspond to the consistency constraint of the local update rule $f_i$.
    Messages $\chi$ are sent back and forth between the nodes.}
    \label{fig:factor-graph}
\end{figure*}

\paragraph{Factorization over the graph.} For the cavity method to be exact, one requires a probability distribution with a tree-like graphical model.
To create such a graphical model for our distribution \eqref{eq:prob_dist}, we need two properties to factorize: The global rule $F$ and the observables~$\Xi$.

First, the constraint on the global rule $F$ factorizes on the local node neighbourhoods as 
\begin{eqnarray}
   \mathbbm{1}\left[F(\xx) = \xx'\right] = \prod_{i=1}^n \mathbbm{1}\left[ f_i\left(x_i,\mathbf{x}_{\partial i}\right)  = x'_i \right]\,,
\end{eqnarray}
which holds since we defined $F$ in terms of the local rules~$f_i$.

We assume that the observables can be factorized similarly, i.e. that we can decompose them as a sum over functions on a single node or edge sequences.
When  $\tx_i = (x_i^1,\cdots,x_i^{p+c})$ is the sequence of states of a single node $i$ in $\txx$ we define the node-localized or edge-localized factorization of an observable as
\begin{eqnarray}
    \Xi(\txx) = \sum_{i\in V} \tilde{\Xi}(\tx_i)\,;\,\,\,\,\,\Xi(\txx) = \sum_{(ij)\in E} \bar{\Xi}(\tx_i,\tx_j).
\end{eqnarray}
where $\tilde{\Xi}_k : S^{p+c} \to \R$ and $\bar{\Xi}_k : S^{p+c} \times S^{p+c} \to \R$.
The application examples in our work require four different observables: The magnetization of the initial configuration $ m_{\textrm{\small init}}$, the average magnetization in the attractor $m_{\textrm{\small attr}}$, { the energy of the configuration after $t$ time steps $e^t$} and for {$c \ge 2$} the fraction of changing nodes (rattlers) in the attractor~$\rho$:
\begin{eqnarray}
    m_{\textrm{\small init}}(\txx) =& \frac{1}{n}\sum_{i\in V} x^1_i \\
    m_{\textrm{\small attr}}(\txx) =& \frac{1}{n}\sum_{i\in V} \frac{1}{c} \sum_{t=p+1}^{p+c}x^t_i\\
    e^t(\txx) =& \frac{1}{m} \sum_{(ij)\in E} x_i^{t} x_j^{t}\\
    \rho(\txx) =&\frac{1}{n}\sum_{i\in V} \mathbbm{1}\left[1 \leq \sum_{t=p+1}^{p+c{-1}} \mathbbm{1}[ x^t_i \neq x^{t+1}_i]\right]
\end{eqnarray}
Each property naturally factorizes either on the nodes or edges.
While we do not consider observables that factorize on local neighbourhoods $(\tx_i,\tx_{\partial i})$, they can be easily integrated into the framework.

Using these factorizations of $F$ and $\Xi_k$, the distribution over sequences $\txx$ from \eqref{eq:prob_dist} can be factorized over the graph to read

\begin{widetext}
\begin{eqnarray*}
    P(\txx) &= \frac{1}{Z} \prod_{i\in V} \underbrace{\left[e^{-\sum_{\tilde{k}}\lambda_{\tilde{k}}\tilde{\Xi}_{\tilde{k}}(\tx_i)} \mathbbm{1}\left[  f_i(x^{(p+c)}_i;\xx^{(p+c)}_{\partial i}) = x^{p+1}_i\right] \prod_{t=1}^{p+c-1}  \mathbbm{1}\left[  f_i(x^t_i;\xx^t_{\partial i}) = x^{t+1}_i\right]\right]}_{\mathcal{A}_i(\tx_i,\txN{i})}\prod_{\{ij\}\in E}\underbrace{\left[ e^{-\sum_{\bar{k}}\lambda_{\bar{k}}\bar{\Xi}_{\bar{k}}(\tx_i,\tx_j)}\right]}_{a(\tx_i,\tx_j)}\,.
\end{eqnarray*}
\end{widetext}
where $\tilde{k}$ and $\bar{k}$ are meant to only sum over the observables that are node and edge localized respectively.
This distribution defines a probabilistic model that can be represented as a factor graph where variables are the local sequences $\tx_i$.
The factors $\mathcal{A}(\tx_i,\txN{i})$ and $a(\tx_i,\tx_j)$ ensure that only $(p/c)$-attractors have a non-zero probability and are biasing towards a given observable if $\lambda_k$ is non-zero.

However, the implicit factor graph is not locally tree-like: If node $i$ and $j$ are connected by an edge, they appear together in the two factors $\mathcal{A}_i, \mathcal{A}_j$. { Hence, for every edge $(i,j)$, there is a loop of length $4$, connecting $x_i \leftrightarrow \mathcal{A}_i \leftrightarrow x_j \leftrightarrow \mathcal{A}_j \leftrightarrow x_i$ (see Appendix Fig.~\ref{fig:naive-factor-graph}). Then, the factor graph is incompatible with an asymptotically exact application of belief propagation.}
Nonetheless, by moving to the edge dual representation of the graph these small loops can be eliminated (the resulting factor graph shows in Fig.~\ref{fig:factor-graph}; for examples of a similar dual construction see e.g. \cite{lokhovDynamicMessagepassingEquations2015,behrensDisAssortativePartitions2022a}).
In the dual space, the variables of the factor graph are tuples of node trajectories $(\tx_i,\tx_j)$ for all $i$ and $j$ that neighbour on the original graph.
\paragraph{BP equations.}As a consequence, the factor graph has the same structure as the original graph.
This leads to BP fixed point equations with messages of the form
\begin{eqnarray}
    \chi_{\tx_i, \tx_j}^{i \to j} \! \! &= \! \frac{1}{Z^{i \to j}} a(\tx_i,\tx_j)\sum_{\txeN{i}{j}} \! \! \mathcal{A}_i(\tx_i,\txN{i})
    \prod_{k \in \partial i \setminus j}\chi_{\tx_k, \tx_i}^{k \to i}. \nonumber
\end{eqnarray}
which may be iterated on a given graph until convergence.
At convergence, the BP result for the free entropy follows as 
\begin{eqnarray}
    n\Phi_{BP} &= \sum_{i \in V} \log(Z^i) - \sum_{(ij) \in E} \log(Z^{ij})\,,\\
    Z^i &= \sum_{\tx_i, \txN{i}} \mathcal{A}_i(\tx_i, \txN{i}) \prod_{j \in \partial i} \chi_{\tx_j, \tx_i}^{j \to i}\,,\\
    Z^{ij} &= \sum_{\tx_i, \tx_j} a(\tx_i,\tx_j)\chi_{\tx_i,\tx_j}^{i \to j} \chi_{\tx_j,\tx_i}^{j \to i}\,.
\end{eqnarray}
We can compute the entropy $s(\xi_1,\dots,\xi_K)$ of the number of valid configurations according to \eqref{eq:configs}, as the constraints in \eqref{eq:constraints} are fulfilled by the fact that we require the BP messages to have converged; they are satisfied at the fixed point.
Note that both the length of the trajectory $p$ and the size of the limit cycle $c$ need to be constant in $n$, as otherwise the limit $n\to\infty$ becomes problematic.

\paragraph{Simplification for random regular graphs.}
The previous equations simplify considerably when we consider regular graphs where all local degrees are $d$.
Furthermore, from hereon we assume that the same local update rule $f: S^{d+1} \to S$ is used for every node and we consider only rules that are independent of the neighbours ordering.
By this permutation symmetry, all BP messages become the same locally as  
\[\chi_{\tx,\ty}^{\to} = \chi_{\tx_i,\tx_j}^{i\to j}\,\,\,\, \forall i,j = 1...n\,,\]
and the BP messages are updated according to
\begin{eqnarray}
    \chi_{\tx,\ty}^{\to} = \frac{1}{Z^{\to}} a(\tx,\ty) \sum_{\tx, \tyy_{[d-1]}} \mathcal{A}(\tx,\tyy_{[d-1]}) \prod_{\ty \in \tyy_{[d-1]}} \chi_{\tx, \ty}^{\to}\,,\label{eq:BP-equation}
\end{eqnarray}
where $\tyy_{[d-1]}$ are the trajectories $(\ty_1,\cdots,\ty_{d-1})$ of the $d-1$ neighbours that are relevant for the local update $f$. 
The free entropy density can be computed as
\begin{eqnarray}
    \Phi_{\mathrm{BP}} &= \log(Z^{\mathrm{fac}}) - \frac{d}{2} \log(Z^{\mathrm{var}})\,,\label{eq:BP-entropy}\\
    Z^{\mathrm{fac}} &= \sum_{\tx, \tyy_{[d]}} \mathcal{A}(\tx,\tyy_{[d]}) \prod_{\ty \in \tyy_{[d]}} \chi_{\tx, \ty}^{\to}\,,\\
    Z^{\mathrm{var}} &= \sum_{\tx, \ty} a(\tx,\ty) \chi_{\ty,\tx}^{\to} \chi_{\tx,\ty}^{\to} \,.
\end{eqnarray}
Eventually, this moves iterating $O(n)$ messages on a full graph to only iterating $O(1)$ messages until convergence.
In addition, for random regular graphs, there are typically no short loops of length $O(\log n)$, which permits the application of BP in the first place.

\paragraph{Solving the equations.}
{For general graphs, the complexity of solving the BDCM equations grow exponentially in $dT=d(p+c)$.}
Similar to the dynamical cavity method it is thus prohibitive to analyse exactly long paths $p$ or large cycles $c$ unless one makes approximations \cite{aurellDynamicMeanfieldCavity2012,zhangInferenceKineticIsing2012,delferraroDynamicMessagepassingApproach2015,barthelMatrixProductAlgorithm2018} or restricts oneself to oriented graphs \cite{neriCavityApproachParallel2009}, graphs with asymmetrically weighted edges \cite{mimuraParallelDynamicsDisordered2009} or unidirectional dynamics with absorbing states {\cite{altarelliOptimizingSpreadDynamics2013,lokhovDynamicMessagepassingEquations2015}}.
In this paper, we will not do any such assumptions or approximations. The problems we address in the next section can be solved using the BDCM directly thanks to the following properties: 
First, for the considered examples the cycle size $c$ is in $\{1,2\}$ \citep{decreasing-energy_functions,derrida1989dynamical}. 
Second, we empirically observe that the dynamics converge in logarithmic time of the system size $n$, so short path lengths $p$ are sufficient to observe interesting properties (see the transient lengths in Fig.~\ref{fig:p-infty-empirics}).
Finally, the local update rules are independent of the order of the neighbourhood which removes the exponential dependence on $d$ via dynamical programming \cite{torrisiUncoveringNonequilibriumStationary2022}.
Overall, we then obtain a time complexity of $O(d 2^{(p+c)})$ per iteration of \eqref{eq:BP-equation}.
Depending on the problem, this allows us to obtain exact results for up to $p \leq  8$ readily. The code for the solver is available at  \href{https://github.com/SPOC-group/backtracking-dynamical-cavity}{github.com/SPOC-group/backtracking-dynamical-cavity}.

\section{Applications}

\subsection{Limiting energy of a quench}

As a first application of BDCM we consider the question posed in the introduction as Example 1, i.e. the limiting energy of the considered dynamics in the antiferromagnetic Ising model.
We note that due to the universality properties discussed e.g. in \cite{zdeborovaConjectureMaximumCut2010a,behrensDisAssortativePartitions2022a} the limiting energy will be the same in the Ising spin glass, i.e. the model with random $\pm 1$ interactions with the zero temperature dynamics. 
For simplicity of the exposition of our results, in what follows, let us describe the case of the antiferromagnet. 

The Ising antiferromagnet on a random regular graph has energy $e(\xx) = \frac{2}{dn}\sum_{(ij) \in E} x_i x_j$. This is the energy that an antiferromagnet at zero temperature is naturally minimizing.
A quench is a zero temperature dynamics where at every time step every spin turns in the direction of its local magnetic field or remains unchanged if the local magnetic field is zero. Initializing spins at random this dynamics starts at zero energy and decreases the energy to a value that we aim to compute. For the antiferromagnet, this corresponds to a specific instance of a local synchronous update rule where each spin (or node) takes the minority state in its neighbourhood.

\begin{table}[ht]
\begin{ruledtabular}
    \begin{tabular}{r|rr|r}
    \textbf{$d=4$}& \multicolumn{2}{l|}{\textbf{BDCM}} & {\textbf{DCM}} \\
    $p$ & $s_p / \log(2)$ & $e_p^*$ & $e'_p$ \\
\colrule
       0 & 0.6026 & -0.3616 & 0.0000\\
1 & 0.8679 & -0.4759 & -0.2812\\
2 & 0.9516 & -0.5156 & -0.4239\\
3 & 0.9812 & -0.5331 & -0.4945\\
4 & 0.9925 & -0.5411 & -0.5261\\
5 & 0.9970 & -0.5447 & -0.5392\\
6 & 0.9988 & -0.5464 & -0.5444\\
7 & 0.9995 & -0.5471 & -0.5463%
       \\\colrule empirical $\tilde{e}_\infty$
        &  \multicolumn{3}{r}{ -0.5475(1)}
       \end{tabular}
    \end{ruledtabular}
\caption{
Normalized entropy $s_p/\log(2)$ of the basins of attraction $p$ steps backward from an attractor obtained from BDCM on the antiferromagnet on a $4$-regular random graph. Column $e^*_p$ gives the energy of the attractor for which this entropy is reached. Column $e'_p$ gives for comparison the energy of the last configuration of a randomly initialized trajectory after $p$ steps obtained with the DCM. 
We compare this with the empirically obtained energy of the attractor from graphs of size $n=10^5$. Table \ref{tab:d=6-energy-comparison} in the appendix provides analogous results for $d=6$. 
}\label{tab:results}
\end{table}

\begin{table}
    \centering
    \begin{ruledtabular}
    \begin{tabular}{l|rrr|rr}
    & \multicolumn{3}{l|}{\textbf{Energy after a quench}} &\multicolumn{2}{l}{\textbf{Equilibrium}} \\
    $d$ &$\tilde{e}_{\infty}$ &  $e_4^*$ & $s_4/\log(2)$ &  $e_{\rm stab}$  & $e_{\rm GS}$ \\
\colrule
4 & -0.5475 & -0.5411 & 0.992 & -0.5774 & -0.7365\\
6 & -0.4764 & -0.4656 & 0.981 & -0.4472 & -0.6097\\
8 & -0.4283 & -0.4151 & 0.969 & -0.3780 & -0.5317\\
10 & -0.3930 & -0.3785 & 0.958 & -0.3333 & -0.4775%
       \end{tabular}
    \end{ruledtabular}
    \caption{We compare the energy to which the synchronous dynamics on $d$-regular graphs with always-stay tie-breaking converges to empirically, $\tilde{e}_{\infty}$ (see Fig.~\ref{fig:AF_convergence} in the appendix) and the energy predicted by the BDCM for path length $p=4$ and the associated entropy.
    This is compared to energies $e_{\rm stab}$ below which equilibrium properties are described with replica symmetry breaking \cite{mezard2001bethe}, and the corresponding 
    ground state energy $e_{\rm GS}$ obtained using the 1-step replica symmetry breaking ansatz computed in  \cite{mezard2001bethe,zdeborovaConjectureMaximumCut2010a}.
    }
    \label{tab:energy-comparison-app}
\end{table}

We now use the BDCM  and compute the size of the basin of attraction after a path of length $p$ of all single-point attractors {($c=1$) that have a given energy $e^*_p := e^{p+1}_p$}.
In Table~\ref{tab:results} we report for every path length $p<8$ the energy $e^*_p$ that maximizes the size of the basin of attraction, i.e. the entropy $e^*_p = \max_e s_p(e)$. We report also the associated maximal entropies $s_p = s(e^*_p)$. 
{ Concretely, the results are obtained as follows: We numerically find a solution of equation \eqref{eq:BP-equation} via fixed point iteration, which in turn gives us the value of $\Phi_{BP}$ in \eqref{eq:BP-entropy}. The value of the energy $e$ is then obtained from \eqref{eq:constraints} using this fixed point. Since the Lagrangian parameter $\xi$ during the fixed point iteration is set to zero, the resulting fixed point will be a local maximum of the entropy and thus the energy is the energy which a typical attractor has. This procedure is equivalent to the maximization of the entropy over the energy but numerically simpler.}

{ To interpret the results,} let us first look at the entropies $s_p$ that quantify how many initial configurations end up in a point-like attractor in $p$ or fewer steps.  
Remarkably, when stepping away only $p=4$ steps backward in time from any point attractor, our analysis of the entropy~$s_p$ shows that one can already reach more than 99\% of the full entropy of the configuration space.
With 3 additional steps, $p=7$, the covered fraction is at more than 99.9\%.
Thus, the size of the basin of attraction for point attractors under this rule quickly encompasses almost all the entropy as $p$ is increased.

We next focus on the value of the energy $e^*_p$ and compare it with the final energy $\tilde{e}_\infty$ obtained numerically on systems of size $n=10^5$. We note that the considered synchronous dynamics converges to an attractor and we thus define the stopping time of the simulation as the time when the attractor is reached. 
We see that $e^*_p$ converges closely to the empirically obtained energy for already very moderate values of $p$. The value of the energy matches in 3 digits after only $p=7$ steps away from the attractor.
At this point, the fraction of the basin of attraction covered amounts up to 99.9\% and provides a nice measure of how close to the limiting $p\to \infty$ result the value $e^*_p$ is.

In Table~\ref{tab:results} we also compare to the   
results of the standard forward DCM \cite{hatchett2004parallel,neriCavityApproachParallel2009,mimuraParallelDynamicsDisordered2009,kanoriaMajorityDynamicsTrees2011,lokhovDynamicMessagepassingEquations2015} for increasing lengths of trajectories $p$. We note that {as far as} we know this energy has not been evaluated before using the DCM and is thus a result of independent interest. We see that the values of the energies also converge to the empirical value very fast, but slightly slower than the BDCM that we propose here. Moreover, the forward DCM does not come up with the natural measure of convergence provided by the value of the entropy~$s_p$. {Overall this example serves us to illustrate the main conceptual differences between the BDCM and DCM in a concise manner.} In the appendix, we report an analogue of Table~\ref{tab:results} for $6$-regular random graphs.

Finally, we comment on the fact that we used the replica symmetric version of the cavity method for the reported results.  { Since we are stepping back from the attractors it could be that describing the statistics of the attractors requires replica symmetry breaking.} 
Following the standard literature on spin glassed on sparse random graphs we analyze the stability towards replica symmetry breaking \cite{mezard1987spin,mezard2001bethe} (more details on the stability in the appendix) to conclude that the reported results are stable where it was possible to check them using the population dynamics method ($p\leq5$ for $d=4,6$ and $p \leq 4$ for $d=8,10$). Also, the empirical results still closely match the ones obtained from the replica symmetric BDCM, see Table \ref{tab:energy-comparison-app}. 
This is quite interesting as in Tab.~\ref{tab:energy-comparison-app} we also give the energy $e_{\rm stab}$ below which replica symmetry breaking (RSB) needs to be considered at equilibrium. 
We obtain $e_{\rm stab}(d=4) = -0.5774$ which is below the energy reached by the quench $\tilde e_\infty$ (as analyzed above). 
However, for $d\ge 6$ the energy reached by a fast synchronous quench (analyzed with the RS approach that is stable towards RSB) is lower than the equilibrium energy at which the RSB need to be taken into account. 
Since the fast quench does not follow equilibrium configurations it thus seems that
it goes to out-of-equilibrium parts of the phase space -- the set of attractors of the randomly initialized quench -- that are replica-symmetric while the majority of configurations at that same energy (the equilibrium) require RSB. This is quite a surprising behaviour, perhaps reminiscent of other problems, such as random graph coloring, where simple algorithms were shown to be finding valid colorings even in regions where the equilibrium is described by replica symmetry breaking \cite{achlioptas2002almost,zdeborova2007phase}. A closer investigation of these replica symmetric sub-spaces of the RSB equilibrium phase is left for future work.

\begin{figure*}[ht]
    \includegraphics[width=\linewidth]{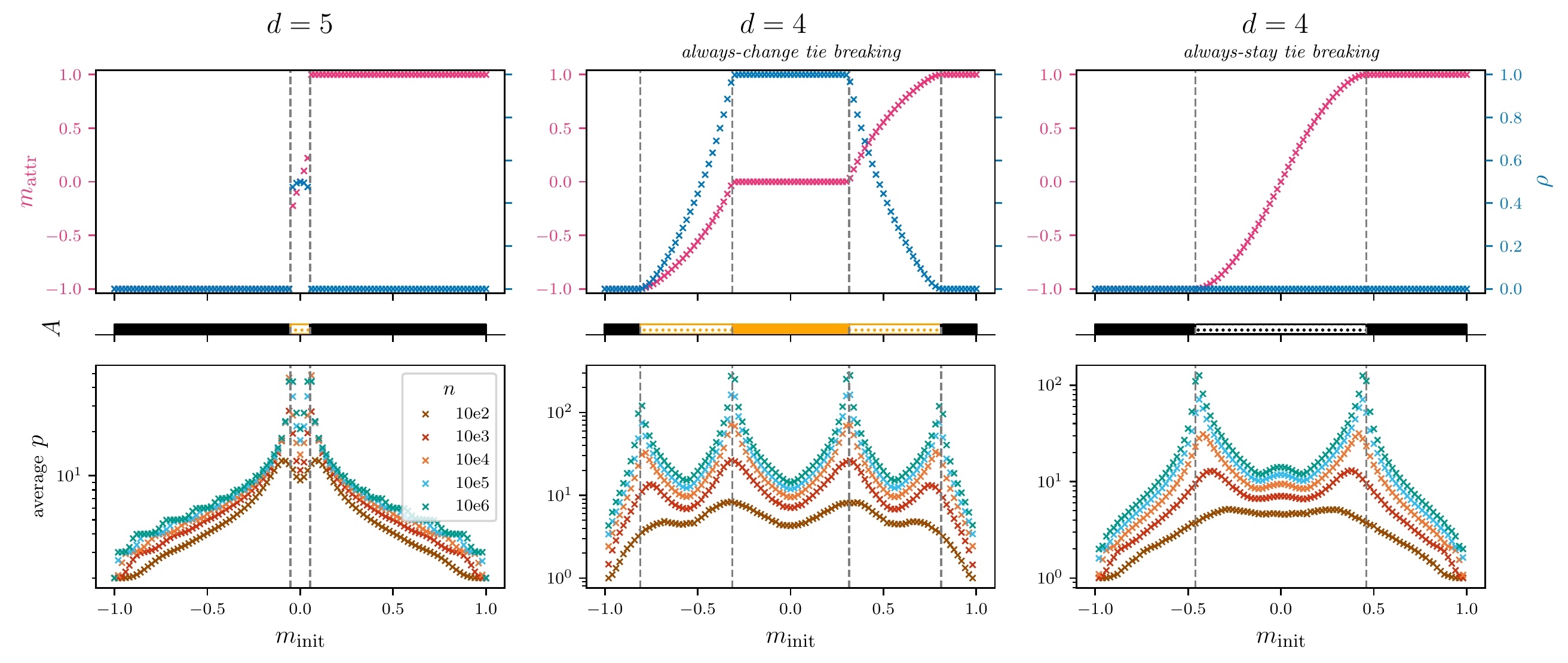}
    \caption{\textbf{Empirics of dynamical phase transitions for majority rules.} 
    We sample trajectories and attractors starting from random initializations with varying magnetization $m_{\textrm{\small init}}$ on instances of random regular graphs of varying sizes $n$.
    On finite systems, every deterministic dynamics reaches a limit cycle in finite time.
        \textit{(Top row)} Properties of the sampled attractors for $n=10^5$: The magnetization of the attractor $m_{\textrm{\small attr}}$ and the fraction of rattlers $\rho$ as a function of $m_{\textrm{\small init}}$.
        \textit{(Middle line)} Combined, the values $m_{\textrm{\small attr}}$ and $\rho$ are linked to the $4$ attractor types of attractors as defined in the main text.
    \textit{(Bottom row)} The average transient length $p$. The positions of the dynamical phase transitions are marked with dashed lines, estimated by the divergencies in the transient lengths (Fig.~\ref{fig:p-transition-emp1},\ref{fig:p-transition-emp2} in the appendix).}
    \label{fig:p-infty-empirics}
\end{figure*}

\begin{figure*}
\centering
\includegraphics[width=\linewidth]{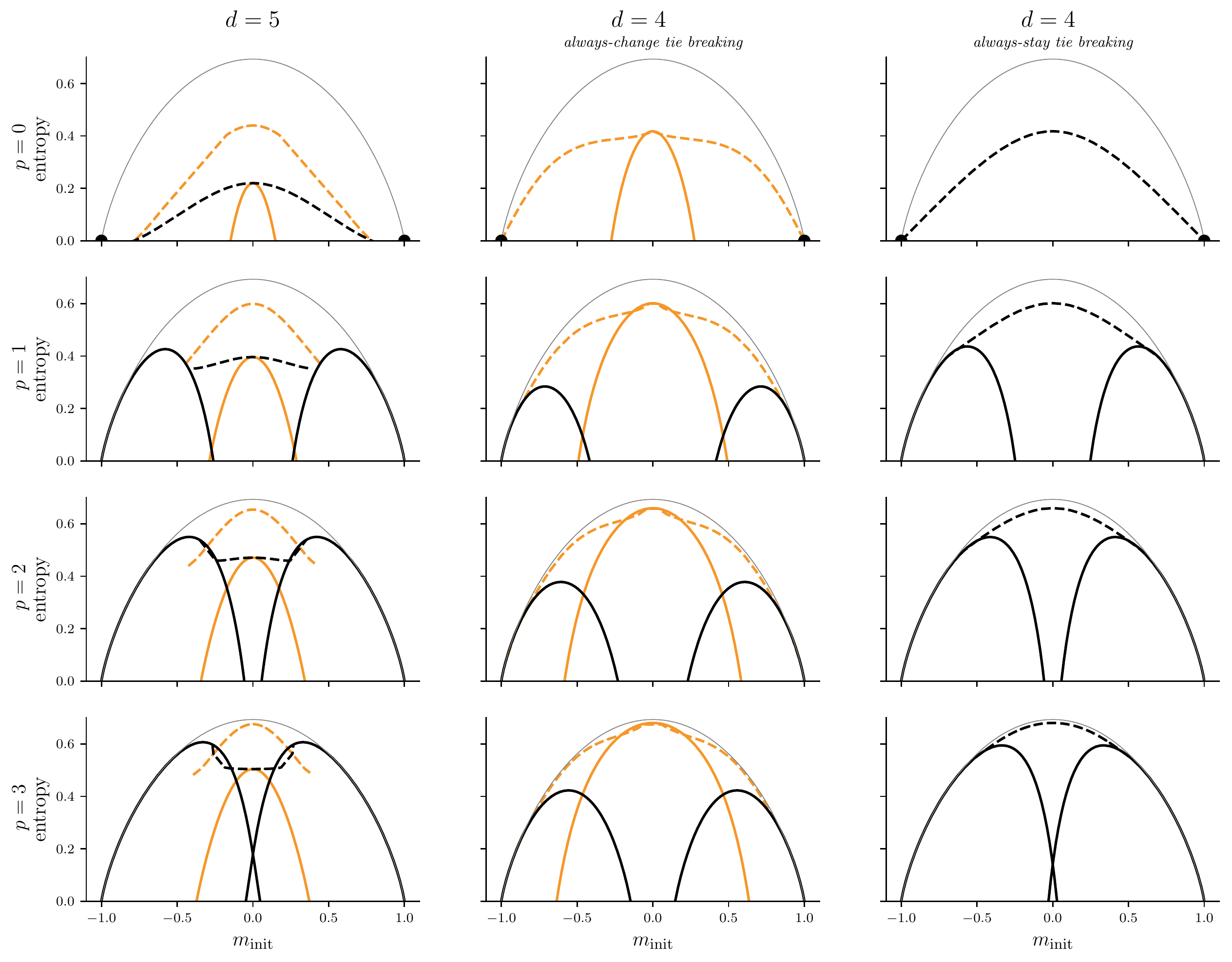}\\
\includegraphics[width=0.9\linewidth]{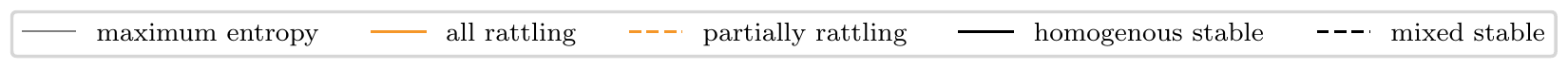}
\caption{\textbf{The BDCM entropy of different attractors and path lengths $p$ for majority rules.} Comparison of the BDCM for on the majority on regular graphs with degree $d=5$ and $4$ with always-stay and always-change tie-breaking. The points in the first row where $p=0$ indicate that the entropy of the homogeneous point attractors is exactly zero at $m_{\textrm{\small init}}=\pm 1$.}
\label{fig:all-overviews-for-different-p}
\end{figure*}

\begin{figure*}
\centering
\includegraphics[width=\linewidth]{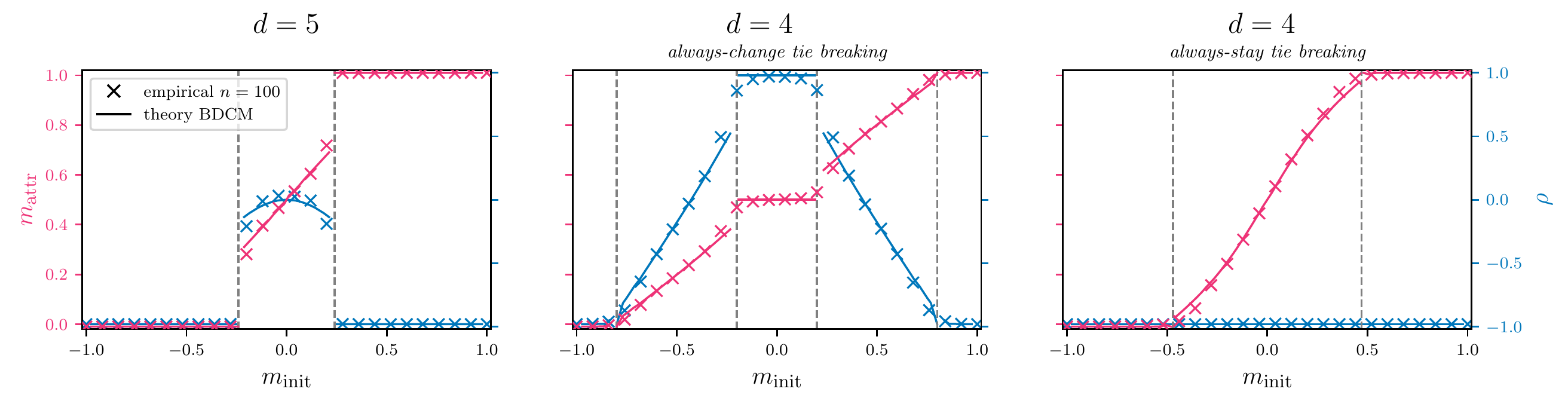}
    \caption{\textbf{Comparison between BDCM and empirics for $p=3$ and majority rules.} For a given $m_{\textrm{\small init}}$ we compare the magnetization and rattlers predicted from the dominating BDCM fixed point, with the respective empirical value. 
    The empirical results were obtained on graphs of size $n=100$, and only backtracking attractors with transients of length $p=3$ were sampled. }\label{fig:p3-theory-vs-empirics}
\end{figure*}

\subsection{Dynamical phase transition for majority rules}

As the second illustration of the BDCM we consider the ferromagnetic Ising model and dynamics corresponding to the majority rule. The questions we investigate here can find applications e.g. in generalized bootstrap-percolation \cite{phase_transitions_in_twoway_bootstrap_percolation}, the zero temperature Glauber dynamics \cite{morrisZerotemperatureGlauberDynamics2011,damronZerotemperatureGlauberDynamics2020}, models of segregation \cite{schellingDynamicModelsSegregation1971}, density classification for cellular automata 
\cite{schonmannBehaviorCellularAutomata1992,busicDensityClassificationInfinite2012}, opinion dynamics \cite{kanoriaMajorityDynamicsTrees2011} or local versions of max or min cut \cite{zdeborovaConjectureMaximumCut2010a, behrensDisAssortativePartitions2022a}.

We consider three basic types of deterministic majority dynamics depending on the degree of the nodes and the type of tie-breaking:
\begin{itemize}
     \item Odd degree, simple \textit{majority rule}: At each time step each spin turns in the direction of the majority of its neighbors. 
     \item Even degree, \textit{always-change tie-breaking} type: Each spin turns to the majority among its neighbors. In the case of balance among the neighbors the spin always changes to the opposite value from the previous time step. 
     \item Even degree, \textit{always-stay tie-breaking} type: Each spin turns to the majority among its neighbors. In the case of balance among the neighbours the spin always remains at the same value as in the previous time step. 
\end{itemize}

We will then investigate the type of attractors to which the dynamics converges when initialized at random but with a magnetization $m_{\textrm{\small init}}$ fixed between $-1$ and $1$. We remind that the majority dynamics always converges to attractors of length either one or two \citep{decreasing-energy_functions,derrida1989dynamical}. We will distinguish between 4 types of attractors, specifically the following ones: 
\begin{itemize}
    \item \includegraphics[width=0.03\linewidth]{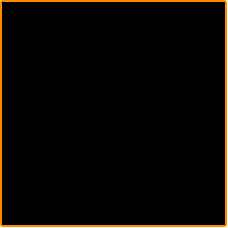} \textit{Homogeneous stable:} These are length 1 attractors with almost all spins either $+1$ or $-1$;\\ $m_{\textrm{\small attr}}\in \{-1,+1\}$ and $\rho=0$.
    \item\includegraphics[width=0.03\linewidth]{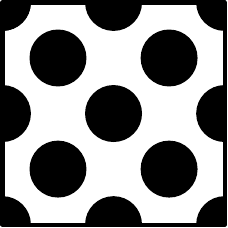}  \textit{Mixed stable:} These are length 1 attractors with a finite fraction of spins in both  $+1$ and $-1$;\\ $m_{\textrm{\small attr}}\in (-1,+1)$ and $\rho=0$.
    \item\includegraphics[width=0.03\linewidth]{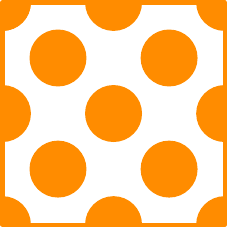} \textit{Partially rattling:} These are length two cycles where a finite fraction of nodes is not changing during the cycle; $\rho \in (0,1)$.
    \item\includegraphics[width=0.03\linewidth]{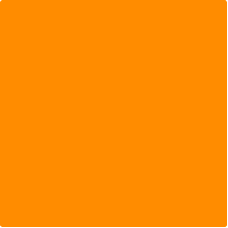} \textit{All rattling:} These are length two attractors where almost all nodes are switching during the cycle; $\rho = 1$.
\end{itemize}
Note that each of the observed attractors falls under exactly one of these four categories.
We emphasize that our definition makes the distinction between $m_{\textrm{\small attr}}$ and $\rho$ only when they correspond to a finite fraction $\Theta(n)$ of the nodes. 
This disregards a subleading number $o(n)$ of nodes that might be of a different sign in a homogeneous stable attractor, or $o(n)$ nodes that are not rattling in the all-rattling attractor.

To make the connection to Example 2 from the introduction, the homogeneous stable attractor corresponds to the all-one configuration. The question is then, what is the least biased value of the initial magnetization so that the dynamics converge with a high probability to such an attractor?

On $d$-regular graphs we can first observe numerically that depending on the initial magnetization $m$ the three dynamical rules converge with high probability to one of the 4 types of attractors defined above, as shown in Fig.~\ref{fig:p-infty-empirics} in the upper panel for degree $d=4$ and $5$. For all three considered dynamical rules, we see that for large enough initial magnetization the dynamics converges to the homogeneous stable attractor. For the simple majority dynamics, odd degree $d$, initial magnetization close enough to zero converges to the partially rattling cycle. For the always-stay tie-breaking dynamics an initial magnetization close enough to zero converges to the mixed stable attractor. For the always-change tie-breaking dynamics initial magnetization close enough to zero converges to the all-rattling cycle, but an intermediate value of magnetization converges to the partially rattling cycle. When we plot the length of the transient to reach the attractor, Fig.~\ref{fig:p-infty-empirics} lower panel for different graph sizes and degree $d=4$ and $5$, we observe logarithmic divergences of the transient lengths at values of the initial magnetization corresponding to those where the type of attractors changes. In statistical physics, a diverging timescale is usually associated with a phase transition, in this case, a \textit{dynamical phase transition}.
Note that simply counting the attractors of various types, as done e.g. in \citep{bray1981metastable,behrensDisAssortativePartitions2022a,hwangNumberLimitCycles2020} does not lead to any sensible explanation of these dynamical phase transitions. 
We will now illustrate how to use the BDCM method to explain and quantify them.

In the BDCM we 
compute the size of the basins of attraction of the various types of attractors or in other words the entropy of the $(p/c)$ backtracking attractors $\txx = (\xx^1,\dots,\xx^{(p+c)})$, conditioned on which type of attractor is expressed in $(\xx^{p+1},\dots,\xx^{p+c})$.
We are able to separate the different types of attractors by introducing $m_{\textrm{\small init}},m_{\textrm{\small attr}}$ and $\rho$ as observables in the BDCM.
We set $c$ according to the attractor length and threshold the $m_{\textrm{\small attr}}$ and $\rho$ to analyze each type separately.
Since we solve the BP equations numerically, this amounts to thresholding the observables $m_{\textrm{\small attr}}$ and $\rho$ with an $\varepsilon=10^{-8}$.
We can isolate the homogeneous and all rattling attractors by conditioning in the BP update on the homogeneous attractors by forbidding all messages $\chi_{\tx\to\ty}$ with $x^{p+1} \neq +1$ with $c=1$ for the all $+1$ and similarly for the all $-1$ attractor.
For the all rattling attractor we similarly forbid any $\chi_{\tx\to\ty}$ with $x^{p+1} = x^{p+2}$ for $c=2$.

Fig.~\ref{fig:all-overviews-for-different-p} depicts the entropy of the basin of attraction for each of the 4 types of attractors for path lengths $p<4$ towards the attractor, in 4 different line types each for one type of attractor, as a function of the initial magnetization. The values of the entropy correspond to the exponent in the number of configurations of magnetization $m_{\textrm{\small init}}$ that converge after $p$ steps to an attractor of the corresponding type. When a line for a given type of attractor is not present it means that this attractor with high probability does not exist for that case.

It is remarkable to note that already with $p=1$ we observe the qualitatively correct picture where the empirically observed attractors indeed correspond to those of the largest entropy. Also, the value of the largest entropy is already relatively close to the total entropy at the corresponding magnetization. The values of the initial magnetization where the maxima change for $p=1$ are of course only rough approximations of those at $p \to \infty$ but the  qualitative behaviour for the three types of dynamics agrees with the one observed empirically. 

For path lengths $p=1,2,3$ we observe that the points where the maximum entropy at that $p$ is reached by a different attractor type are getting closer as $p$ grows to the empirically observed value that would correspond to $p\to \infty$, as reported in Tab.~\ref{tab:transition-majority} in the appendix. The convergence is not at fast as we observed e.g. for the values of the energy in the previous section, but the fact that the maximum entropy converges rather fast to the total entropy indicates the qualitative correctness of the picture.

Next to the values of the entropies, the BDCM also readily provides the values of the attractor magnetization $m_{\rm attr}$ and the fraction of rattlers in the attractor~$\rho$. These values are plotted in Fig.~\ref{fig:p3-theory-vs-empirics} (full lines) for the attractors that correspond to the largest value of the entropy of the basin of attraction after $p=3$ steps backward from the attractor. We observe discontinuities in these parameters at the initial magnetization where the type of attractor changes. 
These data compare qualitatively well with Fig.~\ref{fig:p-infty-empirics} that gives the numerical values for $p\to \infty$. On small graphs, we can also sample very many initial conditions that lead to attractors after $p=3$ steps. Doing so we compare in Fig.~\ref{fig:p3-theory-vs-empirics} with the empirically obtained values of  $m_{\rm attr}$ and $\rho$ observing an excellent quantitative agreement with the theory. The discontinuities are smoothened due to finite-size effects. 

\vspace{-1em}
\section{Conclusion}

In this paper, we introduce the backtracking dynamical cavity method (BDCM) on sparse random graphs for models with synchronous discrete-time deterministic dynamics on discrete variables. We illustrate the method on the problem of computing the limiting energy of a quench, finding cases where the quench goes below the energy that marks the onset of replica symmetry breaking at equilibrium, yet the space of attractors the quench converges to is replica symmetric. We also use the method to characterize dynamical phase transitions occurring as the magnetization of the initial configuration changes in majority-driven dynamics.  

Here we discuss possible extensions and avenues for future work. 
Generalization to dense graphs and continuous variables will require constructing a backtracking version of the dynamical mean-field theory. Such a generalization will open the way to studying limiting dynamics of quenches in dense spin glasses or those of gradient descent in the training of neural networks. 

The effects of replica symmetry breaking can be incorporated straightforwardly following the lines developed in \cite{mezard2001bethe,mezard2003cavity}. Future work will investigate glassy examples where this is relevant.

Another avenue for development is the generalization of the BDCM to stochastically evolving dynamical systems. The dynamical cavity method can be generalized to stochastic dynamics, but more work will be needed to replace the simple counting of states in the basin of attraction with other free-energy-like notions that will be able to pin which of the trajectories are those from random configurations.   

The BDCM shares all the limitations of the usual cavity method \cite{mezard1987spin,mezard2001bethe,mezard2003cavity} in terms of the structure of the interactions that is restricted to mean-field-type of geometries. A clear limitation of the method is the fact that the time $T=O(1)$ and solving the corresponding equations for large values of $T$ becomes cumbersome, yet more work can be done at investigating more efficient solvers for large values of $T$. 
\vspace{-1em}
\begin{acknowledgments}
We thank Guilhem Semerjian, Ginestra Bianconi and Alejandro Lage Castellanos for pointing relevant references to us, { as well as the anonymous referees who provided valuable feedback that improved the quality of the paper. }
Our work was supported by the Czech project AI$\&$Reasoning CZ.02.1.01/0.0/0.0/15\texttt{\char`_}003/0000466 and the European Regional Development Fund, and by SVV-2020-260589. 
\end{acknowledgments}

\newpage

\onecolumngrid
\appendix
\section{Factor graph for the BDCM}

\begin{figure}[h!]
    \centering
 \includegraphics[height=0.22\linewidth]{graph}
    \includegraphics[width=0.58\textwidth]{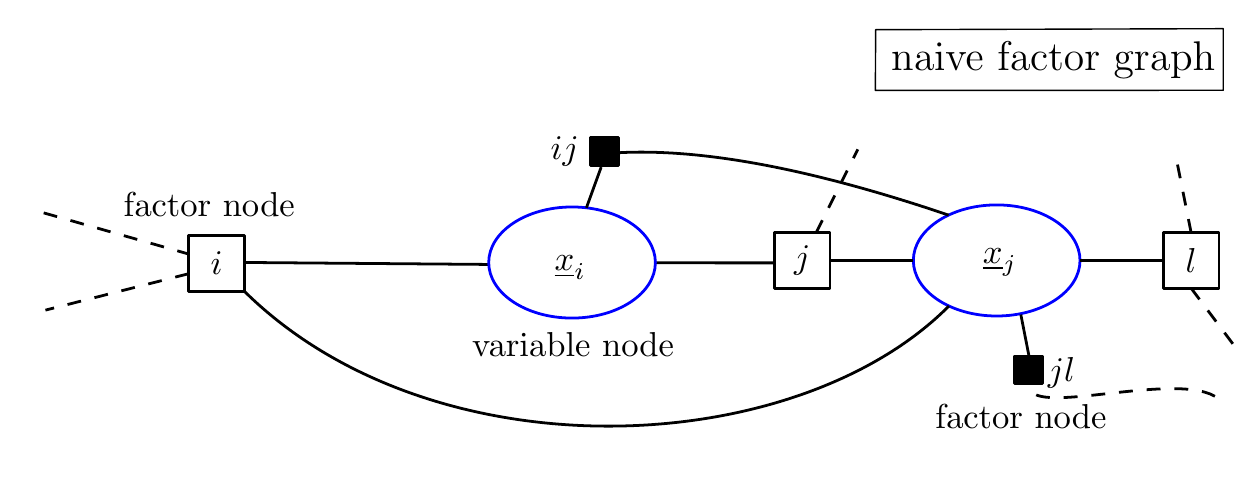}
    \caption{{ Naive construction of a factor graph, which only contains a single node trajectory $\tx_i$ in each variable node. This leads to loops of length $4$, a problem for applying BP.}}
    \label{fig:naive-factor-graph}
\end{figure}

{ In Fig.~\ref{fig:factor-graph} the factor graph with dual variables $(\tx_i,\tx_j)$ is shown, which is subsequently used to derive the BDCM equations.
This dual view allows one to prevent short loops in the factor graph; when the naive construction is used where a variable node contains exactly one nodes trajectory $\tx_i$, both the factor nodes $i$ (representing the update constraint on node $i$ and its neighborhood) and the factor nodes $ij$ (representing the observables on edges) need to be connected to the relevant variables.
As shown in Fig.~\ref{fig:naive-factor-graph}, this leads to short loops of length $4$, which can be prevented by using the dual representation in Fig.~\ref{fig:factor-graph}.}

\section{Stability of the BDCM fixed points towards RSB}

Since the replica symmetric ansatz (RS) that we follow in this paper may not be correct, we check whether the fixed points obtained from the BDCM in our results are stable towards replica symmetry breaking via population dynamics and its convergence analysis.

We initialize the population dynamics with 300 BP messages initialized Gaussian i.i.d. and then normalized to~$1$.
In every iteration, $80\%$ of the messages are updated according to \eqref{eq:BP-equation} applied to $d-1$ randomly selected neighbours.
This process is run until convergence. 
We then check whether the distribution concentrates on a delta function identical to the fixed point found from the RS, or whether it converged to a non-delta distribution over messages.
If the first is the case, we say the distribution is stable to RSB, otherwise, it is unstable.

For the experiments from Tab.~\ref{tab:results} we validate that the fixed points are indeed stable for $p\leq 5$ and for Tab.~\ref{tab:energy-comparison-app} for $p\leq4$.
Since for the larger $p$ the computational time is exceedingly large, we were not able to verify these results.
For the BDCM fixed points used in Fig.~\ref{fig:all-overviews-for-different-p} we check the stability of the dominating fixed point in each of 80 equally sized intervals we divided $m_{\textrm{\small init}}\in \{-1,+1\}$ into, for $p=0,1,2,3$.

\section{Additional results and supporting empirics}

\subsection{Limiting energy of a quench}

\paragraph{Energetic results for $d=6$.}
In Tab.~\ref{tab:d=6-energy-comparison} we present the BDCM and DCM results for the energy of the antiferromagnet or a spin glass on a $6$-regular random graph, analogous to Tab.~\ref{tab:results} from the main text.
As before, for large enough $p$ the BDCM and DCM are close to the energy that is observed empirically.
Again, the BDCM is slightly more precise given the same number of steps into the attractor's basin $p$.
The size of the basin of attraction as measured by the normalized entropy $s_p/\log(2)$ converges very fast to 1 but slightly slower in $p$ than for $d=4$.
{ For the interested reader, we also provide the energies measured by the BDCM at the start of the path going into the attractor, i.e. the energy $e^{\mathrm{start}}$. The observation that the $e^{\mathrm{start}}$ quickly grows to zero, the energy of a random configuration, implies that it only takes a logarithmic number of steps between the inital random configuration and the final energy. This can be viewed as an alternative measure to the entropy, that allows one to assess the quality of the BDCM prediction after $p$ steps.}

\begin{figure*}[h]
    \centering
    \includegraphics[width=0.24\linewidth]{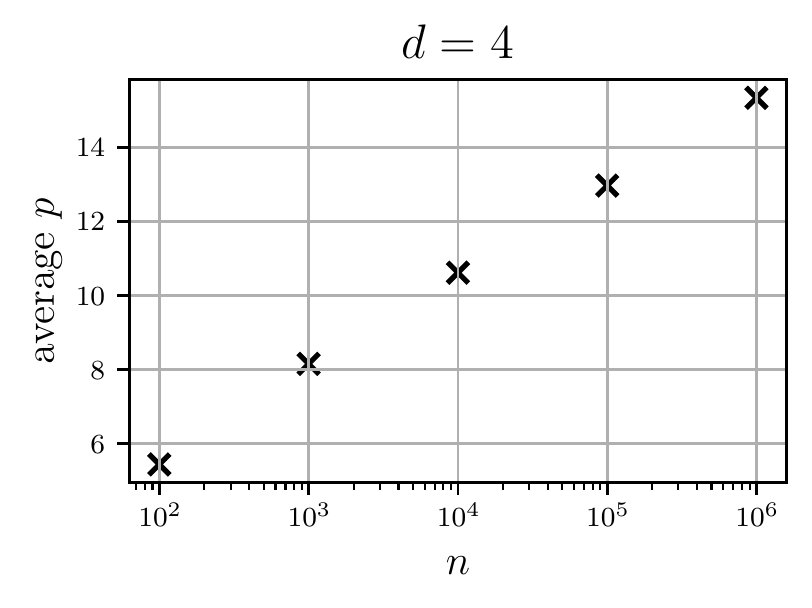}
    \includegraphics[width=0.24\linewidth]{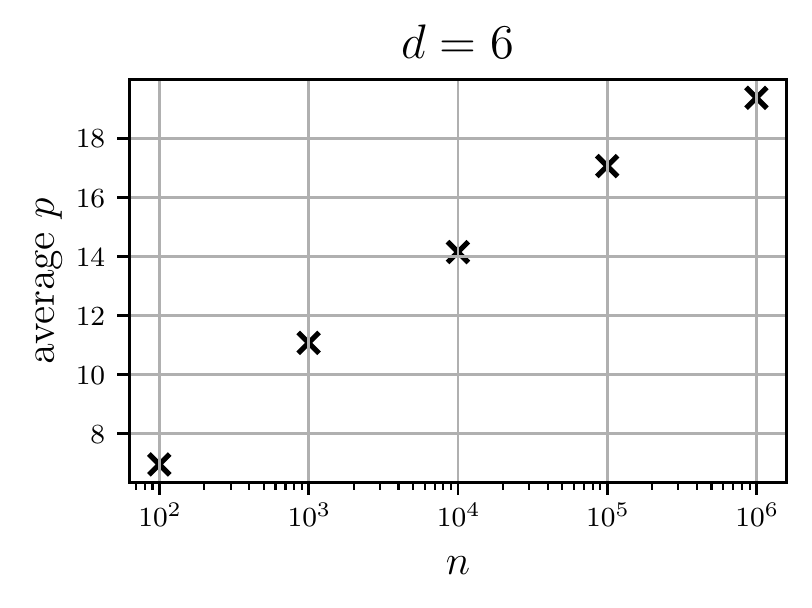}
    \includegraphics[width=0.24\linewidth]{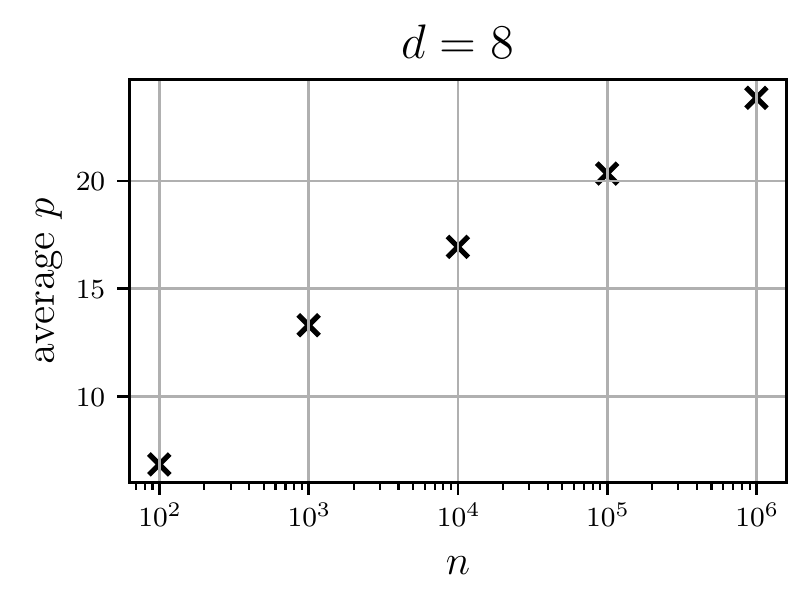}
    \includegraphics[width=0.24\linewidth]{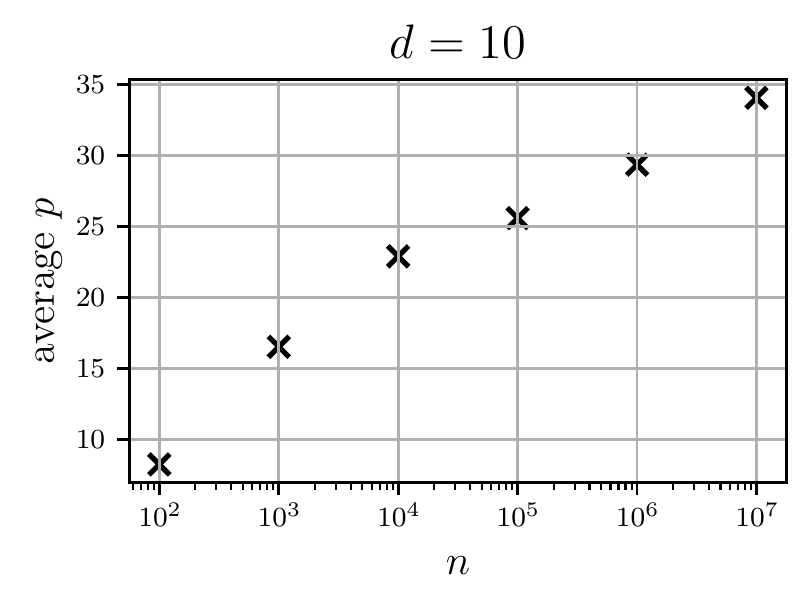}
    \\
    \includegraphics[width=0.24\linewidth]{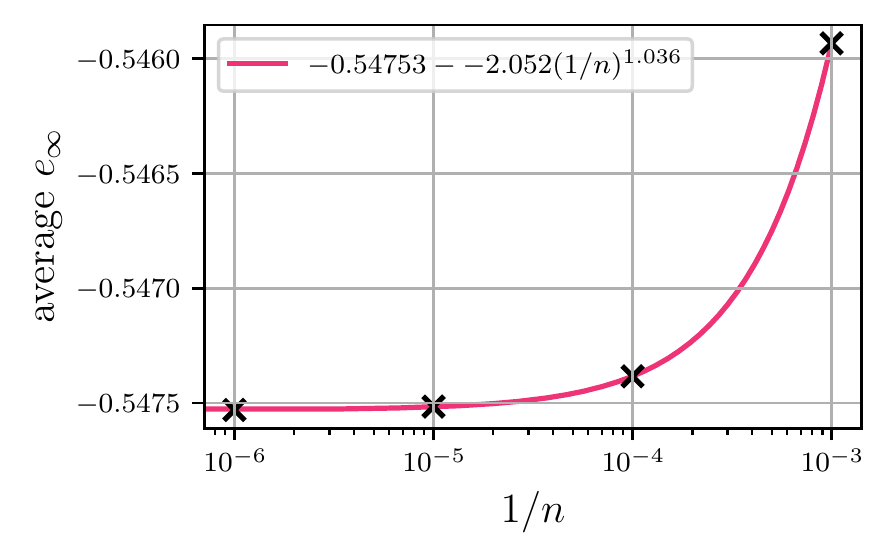}
    \includegraphics[width=0.24\linewidth]{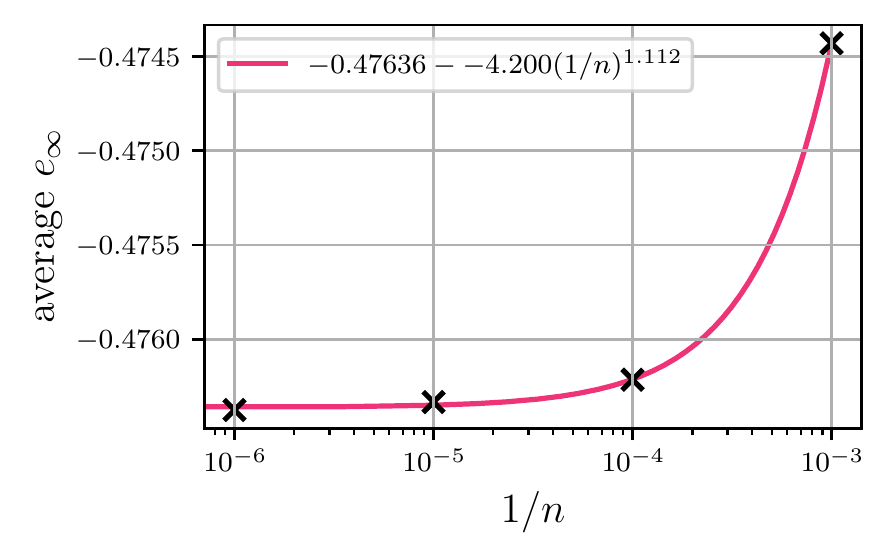}
    \includegraphics[width=0.24\linewidth]{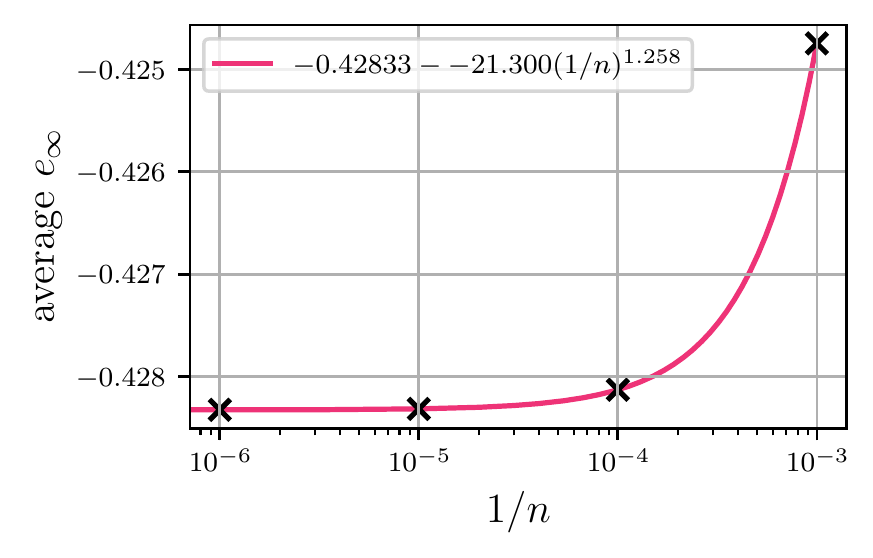}
    \includegraphics[width=0.24\linewidth]{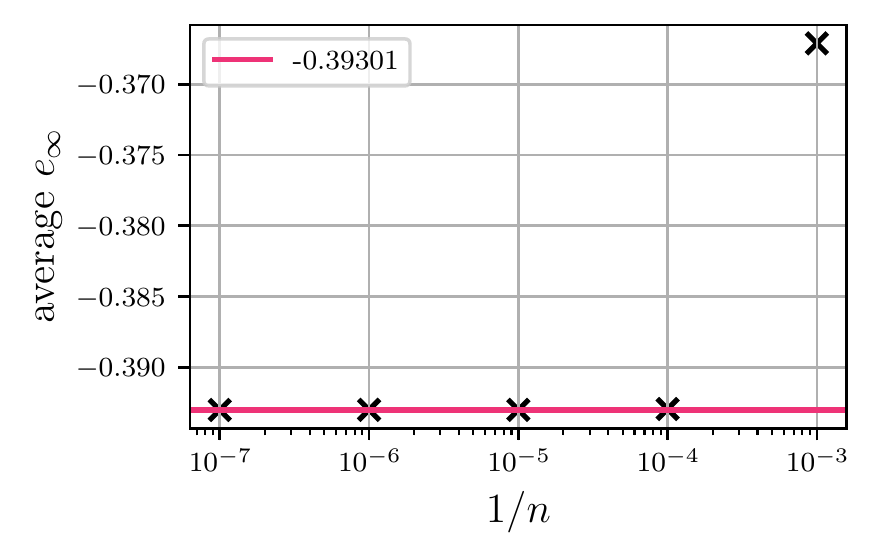}
    \caption{\textbf{Empirical dynamics of the quench.} Results for the energy reached from a randomly sampled balanced initial configuration on $d$-regular graphs with $n$ nodes. \textit{(Top row)} Average transient length until an attractor is reached. We see a growth logarithmic in $n$. \textit{(Bottom row)} Average energy in the attractor for different $n$. The red line extrapolates the energy for $n \to \infty$. }
    \label{fig:AF_convergence}
\end{figure*}

\paragraph{Empirical results.}
To determine the limiting energy of the quench  empirically, we sample $2,048$ random regular graphs per graph size $n$.
We initialize them with a random configuration where the number of $+1$ and $-1$ spins is equal.
The synchronous dynamics are run until convergence and we report the sample average for the energy in the first configuration of the attractor.
These empirical results for $p \to \infty$ are shown in Fig.~\ref{fig:AF_convergence}.
We extrapolate $\tilde{e}_\infty$ in terms of $n$; this is used to compare against the BDCM and DCM results in Tab.~\ref{tab:results} and \ref{tab:d=6-energy-comparison}.

\paragraph{The energetic landscape of the spin glass.}
In Tab.~\ref{tab:energy-comparison-app} we compare the energies reached by the quench and obtained from the BDCM with characteristics of the equilibrium energetic landscape.
Note, that the energy reached by the synchronous dynamics for $d \geq 6$ is lower than the energy below which replica symmetric breaking is required to describe the equilibrium $e_{\rm stab}$. However, the BDCM is still stable towards RSB in the investigated cases of $d=6$ and $p \le 5$ and $d=8,10$ with $p \leq 4$. For larger values the stability check we performed is numerically too costly, we anticipate we would also find stable results.  
 
\begin{table}[h]
    \begin{ruledtabular}
    \begin{tabular}{r|rrr|r}
    \textbf{$d=6$}& \multicolumn{3}{l|}{\textbf{BDCM}} & {\textbf{DCM}} \\
    $p$ & $s_p / \log(2)$ & $e_p^*$ & {$e^{\mathrm{start}}_p$} & $e'_p$ \\
\colrule
0 & 0.5542 & -0.3138 & -0.3138 & 0.0000\\
1 & 0.8223 & -0.4079 & -0.1520 & -0.1953\\
2 & 0.9205 & -0.4407 & -0.0865 & -0.3100\\
3 & 0.9617 & -0.4568 & -0.0531 & -0.3803\\
4 & 0.9807 & -0.4656 & -0.0337 & -0.4226\\
5 & 0.9901 & -0.4705 & -0.0216 & -0.4477\\
6 & 0.9949 & -0.4732 & -0.0138 & -0.4617%
       \\\colrule empirical $\tilde{e}_\infty$
        &  \multicolumn{4}{r}{ -0.4764(1)}
       \end{tabular}
    \end{ruledtabular}
    
    \caption{
    Same as Table~\ref{tab:results} for random regular graphs of degree $d=6${, except that additionally $e^{\mathrm{start}}_p$ is provided, the energy at the start of the path for the BDCM}.}  
    \label{tab:d=6-energy-comparison}
\end{table}

\newpage

\subsection{Dynamical phase transitions for majority rules}

In Fig.~\ref{fig:p-infty-empirics} and \ref{fig:all-overviews-for-different-p} we find four different dynamical phase transitions. 
By the $\pm 1$ symmetry we only look at the transitions occurring for $m_{\textrm{\small init}}>0$.
These are 
\begin{itemize}
    \item[(1)]  \includegraphics[width=0.015\linewidth]{partial0rattler02.pdf} $\leftrightarrow$ 
    \includegraphics[width=0.015\linewidth]{homogenous02.pdf} for $d=5$
    \item[(2)] \includegraphics[width=0.015\linewidth]{all0rattler02.pdf} $\leftrightarrow$ \includegraphics[width=0.015\linewidth]{partial0rattler02.pdf} for $d=4$, always-change tie-brekaing
    \item[(3)] \includegraphics[width=0.015\linewidth]{partial0rattler02.pdf} $\leftrightarrow$ \includegraphics[width=0.015\linewidth]{homogenous02.pdf}  for $d=4$, always-change tie-brekaing
    \item[(4)] \includegraphics[width=0.015\linewidth]{mixed0stable02.pdf} $\leftrightarrow$ \includegraphics[width=0.015\linewidth]{homogenous02.pdf}  for $d=4$, always-stay tie-breaking
\end{itemize}
Note that bounds on the transition (1) as well as its empirical positions were investigated thoroughly for a range of different $d$ in \cite{kanoriaMajorityDynamicsTrees2011}.

\begin{figure*}[h]
    \centering
    \includegraphics[width=0.24\linewidth]{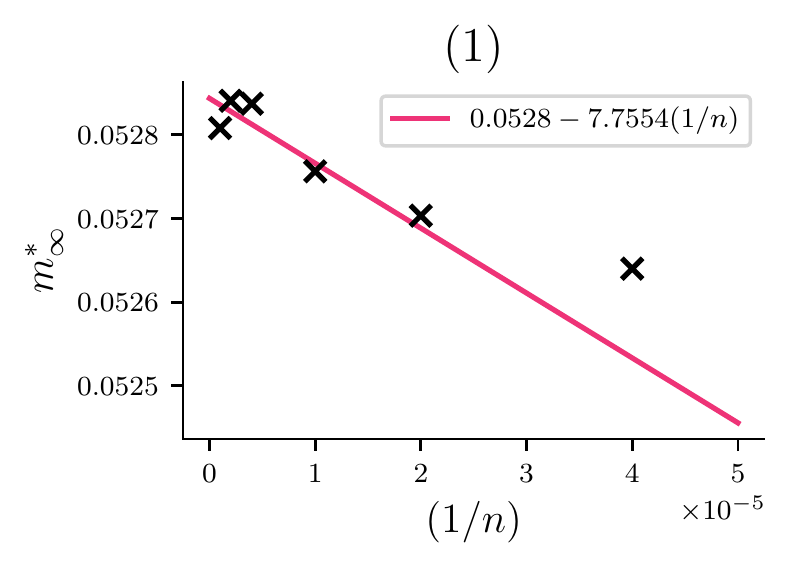}
    \includegraphics[width=0.24\linewidth]{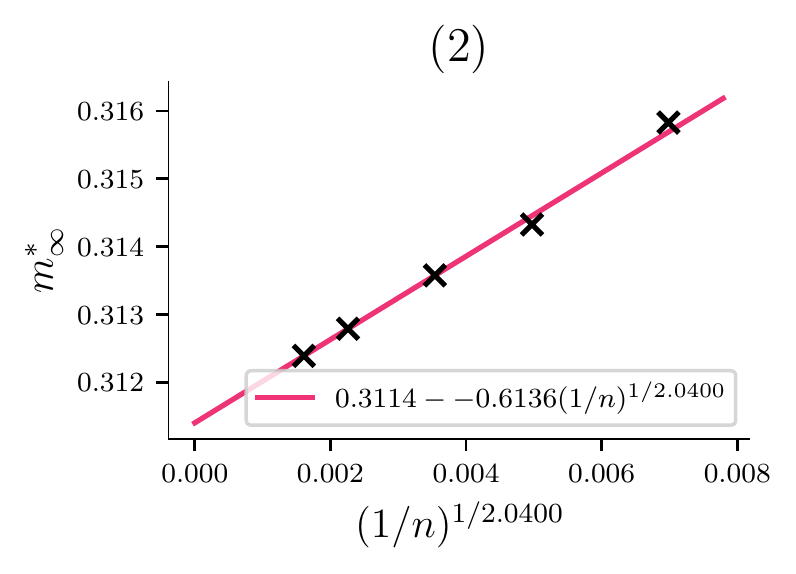}
    \includegraphics[width=0.24\linewidth]{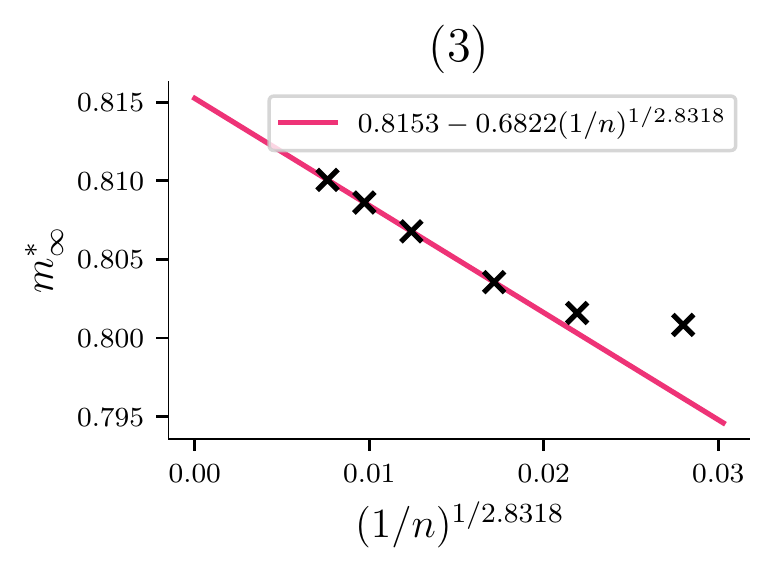}
    \includegraphics[width=0.24\linewidth]{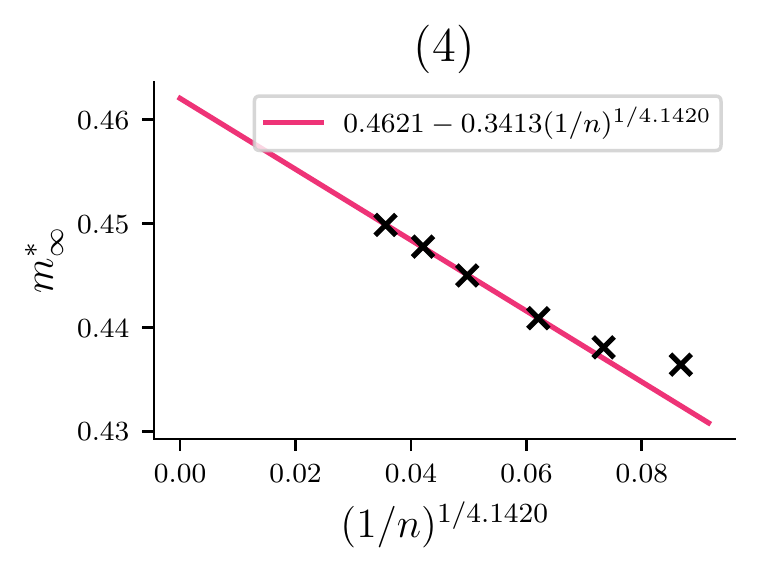}
    \caption{\textbf{Transient lengths:} The average transient length maximized over the magnetization $m_{\rm init}$ to determine the position of the dynamical phase transitions. We show the four types of dynamical phase transitions from Tab.~\ref{tab:transition-majority}. The position $\tilde{m}^*_\infty(n)$ of the maximal average transient spike is shown and extrapolated to $n\to \infty$. We use these results as a reference for the empirical phase transitions at $p \to \infty$. }
    \label{fig:p-transition-emp1}
\end{figure*}

\begin{figure*}[h]
    \centering
    \includegraphics[width=0.24\linewidth]{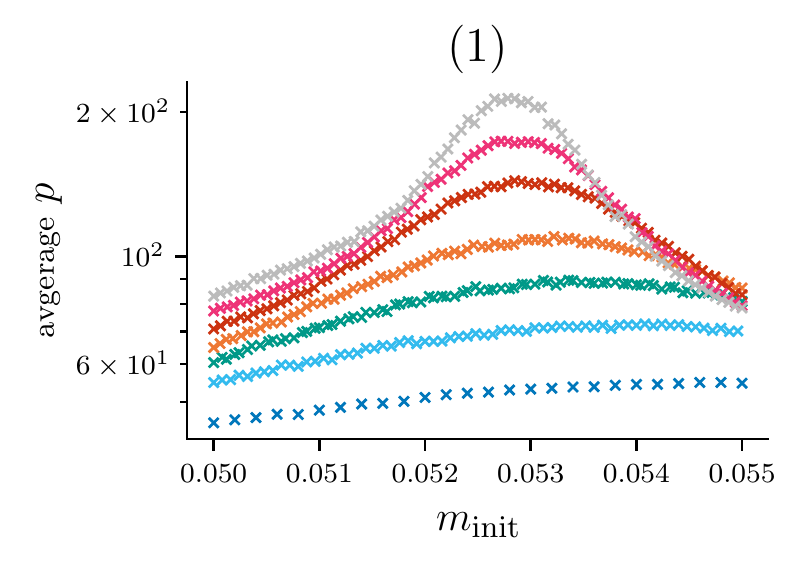}
    \includegraphics[width=0.24\linewidth]{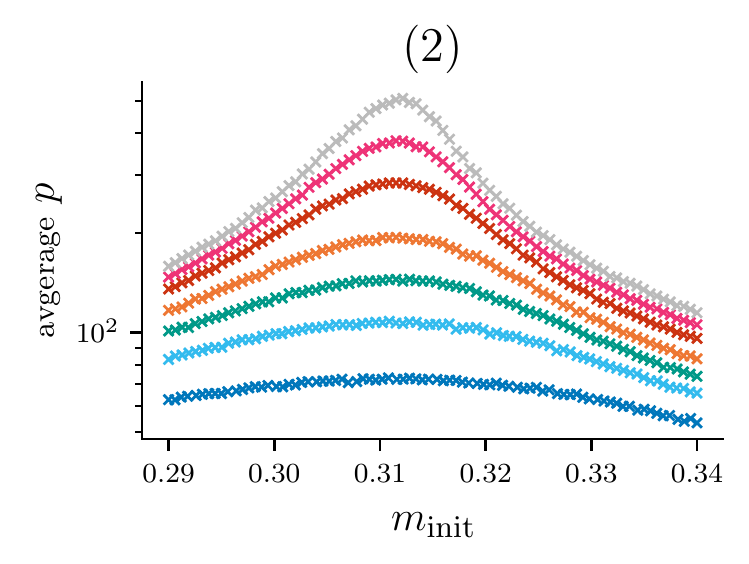}
    \includegraphics[width=0.24\linewidth]{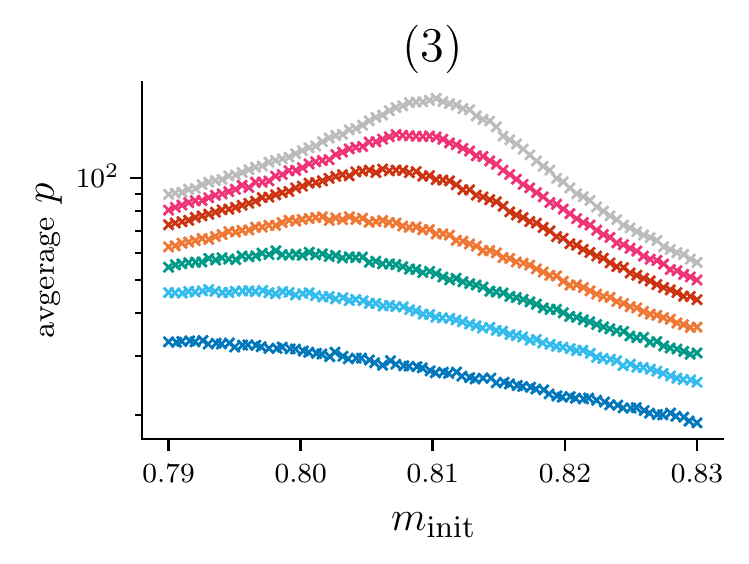}
    \includegraphics[width=0.24\linewidth]{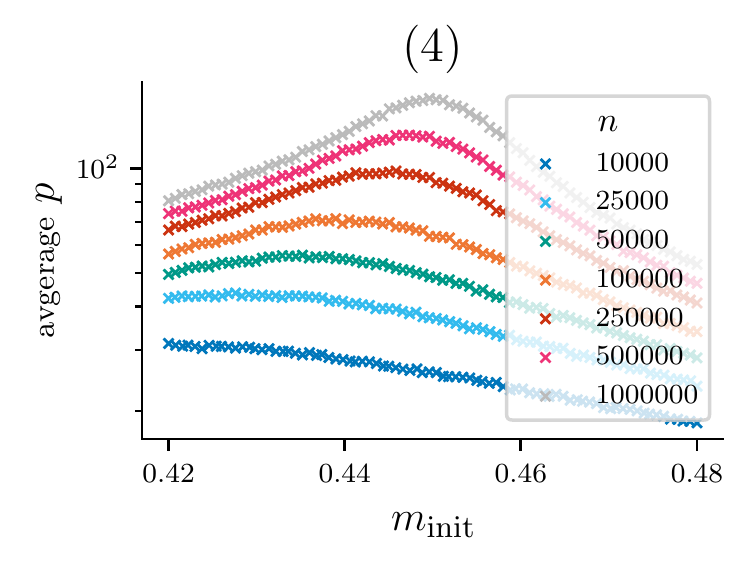}
    \caption{\textbf{Zoom in on the transient lengths for different dynamical phase transitions.} According to the four types of transitions defined in Tab.\ref{tab:transition-majority}, this plot shows zoom-ins on the transients obtained in the same manner as in the lower panel of Fig.~\ref{fig:p-infty-empirics}. Every sample is the average over the dynamics run on 2,048 graph instances with random initializations.}
    \label{fig:p-transition-emp2}
\end{figure*}

\paragraph{Empirical results.}
We obtain the accurate positions of the dynamical phase transitions above for $p \to \infty$ by numerical simulations.
Fig.~\ref{fig:p-transition-emp2} shows a zoom-in for the lower panel of Fig.~\ref{fig:p-infty-empirics}; the averages of the transient lengths $p$ as a function $m_{\textrm{\small init}}$ for different sizes $n$. Averages are taken over $4,096$ samples of random regular graphs and initial configurations.
Then, Fig.~\ref{fig:p-transition-emp1} shows the extrapolation of the position of the maxima for large $n$.

\paragraph{Transitions from the BDCM.}
We give the locations of dynamical phase transitions shown in Fig.~\ref{fig:all-overviews-for-different-p} in Tab.~\ref{tab:transition-majority}.
The values of the transitions for small values of $p$ with the BDCM are not very close to the empirically found transitions $\tilde{m}_\infty^*$.
However, as $p$ grows they become more accurate.
This is in line with the observation that for the example of the quench, the fraction of the basin of attraction was much closer to one than it is in the examples shown here.

\newcommand{\negphantom}[1]{\settowidth{\dimen0}{#1}\hspace*{-\dimen0}}

\begin{table}
 \begin{ruledtabular}
    \begin{tabular}{l|rr|rr|rr|rr}
    & \multicolumn{2}{l|}{$d=5$} & \multicolumn{4}{l|}{$d=4$ \textit{always-change}} & \multicolumn{2}{l}{$d=4$ \textit{always-stay}}\\
    &\multicolumn{2}{l|}{ (1)
    \includegraphics[width=0.015\linewidth]{partial0rattler02.pdf} $\leftrightarrow$ 
    \includegraphics[width=0.015\linewidth]{homogenous02.pdf} } &  \multicolumn{2}{l|}{  (2) \includegraphics[width=0.015\linewidth]{all0rattler02.pdf} $\leftrightarrow$\includegraphics[width=0.015\linewidth]{partial0rattler02.pdf}}
    &  \multicolumn{2}{l|}{(3) \includegraphics[width=0.015\linewidth]{partial0rattler02.pdf} $\leftrightarrow$ \includegraphics[width=0.015\linewidth]{homogenous02.pdf} }
    & \multicolumn{2}{l}{ (4) \includegraphics[width=0.015\linewidth]{mixed0stable02.pdf} $\leftrightarrow$ \includegraphics[width=0.015\linewidth]{homogenous02.pdf}} \\
    $p$ &  $m_{p}^*$ & $s(m_{p}^*)/H(m_{p}^*)$  &  $m_{p}^*$ & $s(m_{p}^*)/H(m_{p}^*)$   &  $m_{p}^*$ & $s(m_{p}^*)/H(m_{p}^*)$   &  $m_{p}^*$ & $s(m_{p}^*)/H(m_{p}^*)$   \\
    \colrule
    1 & 0.443 & 0.627 & 0.132 & 0.832 & 0.902 & 0.925 & 0.617 & 0.880\\
2 & 0.302 & 0.795 & 0.179 & 0.912 & 0.872 & 0.968 & 0.496 & 0.951\\
3 & 0.231 & 0.874 & 0.200 & 0.945 & 0.855 & 0.983 & 0.457 & 0.977\\
    \colrule
    $\tilde{m}^*_\infty$ & 0.0528(1)\negphantom{8(1)}&   & 0.312(1)\negphantom{(1)}  &   & 0.81(1)\hspace{-0.8em}  &   & 0.46(1)\hspace{-0.8em}  &   
       \end{tabular}
    \end{ruledtabular}
\caption{\textbf{Dynamical phase transition for fixed $p$ via BDCM.}
    The table shows all of the different types of dynamical phase transitions that are observed in Fig.~\ref{fig:all-overviews-for-different-p} (considering the $\pm 1$ symmetry).
    Four different dynamical phase transitions occur for the majority rules between different types of attractors: 
     \protect\includegraphics[width=6pt]{homogenous02.pdf} \textit{homogenous stable}, \protect\includegraphics[width=6pt]{mixed0stable02.pdf}  \textit{mixed stable}, \protect\includegraphics[width=6pt]{partial0rattler02.pdf} \textit{partially rattling} and \protect\includegraphics[width=6pt]{all0rattler02.pdf} \textit{all rattling}.
     In addition, we show the size of the basin of attraction taken into account proportional to the maximal entropy $H(m)$ for a configuration of a given magnetization $m$.
     We compare the results to the empirically observed transition at $p \to \infty$
    (from Fig.~\ref{fig:p-transition-emp1}).
    }
    \label{tab:transition-majority}
\end{table}

\newpage

\end{document}